\documentclass{elsart}
\usepackage[english]{babel}
\usepackage[dvips]{graphicx}

\begin{document}
\bibliographystyle{unsrt}

\begin{frontmatter}
\title{Mechanism of epitaxial self-assembly of Fe nanowedge islands on Mo(110)}

\author{I. V. Shvets\thanksref{correspond}},
\author{S. Murphy},
\author{N. Berdunov},
\author{V. Usov},
\author{G. Mariotto}

\address{SFI Nanoscience Laboratory, Physics Department, Trinity College, Dublin 2, Ireland}

\thanks[correspond]{Corresponding author: Fax: (+)353 1 6083228, E-mail: ivchvets@tcd.ie}

\author{S. Makarov}

\address{Allegro Technologies Ltd., Unit 8, Enterprise Centre, Pearse St., Dublin 2, Ireland}

\begin{abstract}

The deposition of Fe films in a nominal thickness range of \mbox{$2.5 \leq \textrm{d} \leq 8$ \AA} on the Mo(110) 
surface at elevated temperatures results in the formation of distinctive nanowedge islands supported on a pseudomorphic 
Fe layer. We propose a model explaining the growth mode of these wedge-shaped Fe islands. The model is based on the 
strain produced in the substrate around each island, by the lattice mismatch between the film and substrate. Fe adatoms 
migrate towards the islands due to the influence of this strain, which is related to the thickness and size of each 
island. The adatoms subsequently enter the islands not only through their thin ends where the island can be only two 
monolayers thick but on the contrary, through a vertical climb along the sides of the thicker end, which can be tens of 
layers thick. This mode of mass transport is again driven by strain, corresponding to the energy reduction through 
movement towards the location where the interatomic spacing in the island corresponds to the bulk value. A key element 
of the model is that misfit dislocations are formed in the lower layers of the nanowedge, which act as migration 
channels for the vertical climb.

\end{abstract}

\end{frontmatter}

\section{Introduction}

\begin{figure}
\center
\includegraphics[width=0.5\textwidth,height=0.35\textheight]{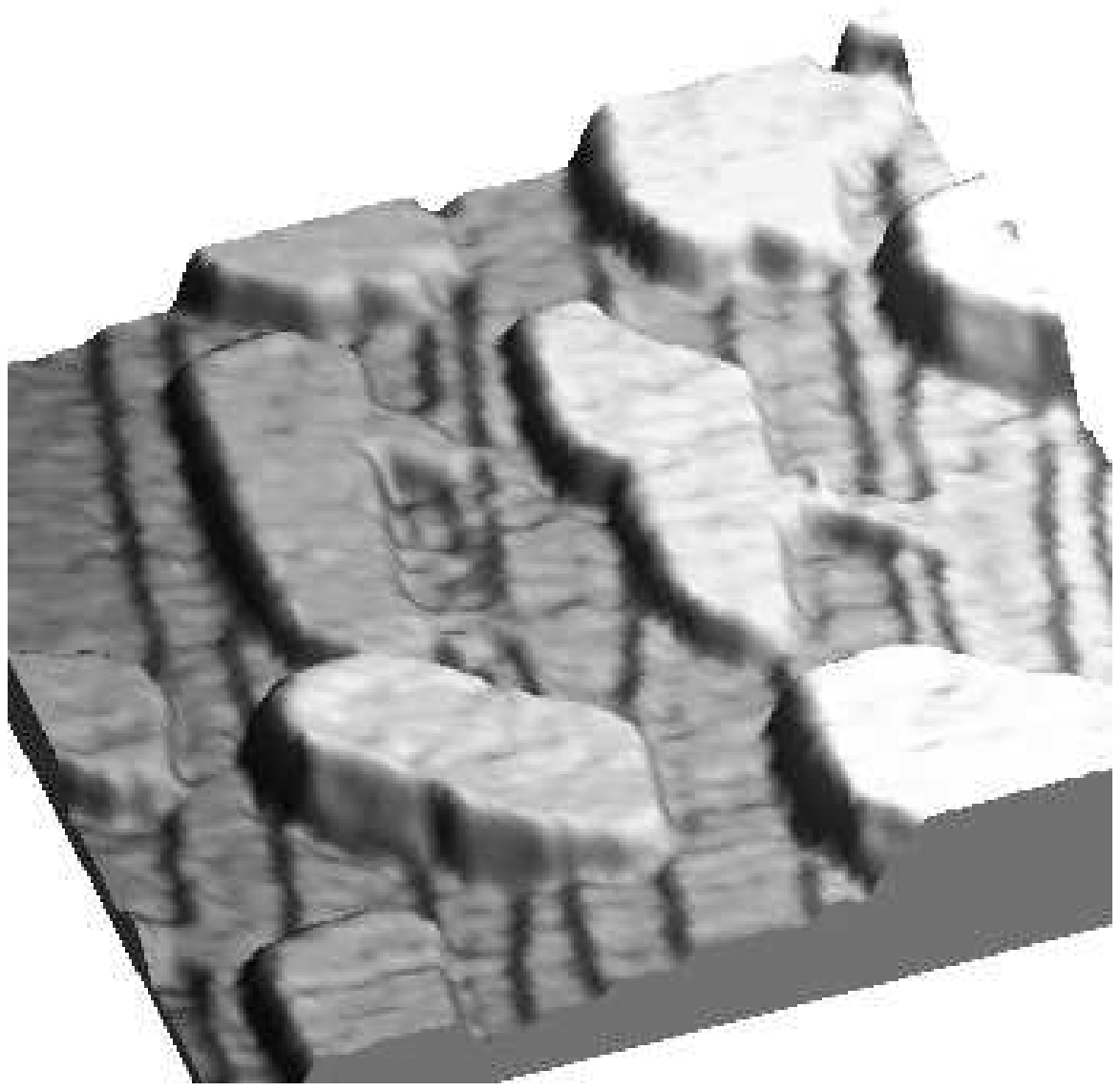}
\caption{Nanowedge islands typically formed during Fe growth on Mo(110) at elevated temperatures. The nanowedges are 
formed on top of a closed pseudomorphic Fe layer.}
\label{figure01}
\end{figure}

The concept of nanopatterning of surfaces through self-assembly of heteroepitaxially grown nanostructures has received 
significant attention in recent years with regard to meeting the increasing demands for miniaturization in high-tech 
applications. Examples are to be found in metal \cite{bru98}, semiconductor \cite{bru02,tei02} and insulator 
\cite{ale98,bat01,vas03} heteroepitaxy, where a variety of nanostructures such as nanowires, pyramids or hut-shaped 
islands can be formed. In many cases, the self-assembly process is underpinned by the strain relaxation behaviour within 
the system. We consider the case for the Fe/Mo(110) epitaxial system, where there is a large lattice mismatch of 
(a$_{Fe}$ - a$_{Mo}$)/a$_{Mo}$ = -8.9 \%. In earlier studies of epitaxial Fe growth on Mo(110) \cite{mal98,mur02}, it 
was demonstrated that under certain conditions the deposited films are characterised by the formation of arrays of 
nanoscale wedge-shaped islands that range in thickness from one or two atomic layers at their thin end to many layers 
thickness at the other (\mbox{Fig. \ref{figure01}}). These nanowedges always grow in such a way as to maintain a 
continuous (110) surface that is unbroken by steps. Similar nanowedge islands have been obtained by depositing Fe, Cu, 
Ni and Ho on W(110) substrates \cite{bet95,res99,sch98,pia00} and Pb on Si(111) \cite{alt97}. It is likely that under 
the correct growth conditions this growth mode can also be obtained in other epitaxial systems. So far, no significant 
attempt has been made to understand the reason for this growth mode. In this study, we explore the mechanisms guiding 
the nanowedge formation in the Fe/Mo(110) epitaxial system and show how they are linked to the mismatch-induced strain. 

\section{Experimental}

The experimental results were obtained on two different Mo(110) surfaces. The first had a miscut of 
\mbox{0.65$^{\circ}$} from the (110) plane, corresponding to an average terrace width of \mbox{$\sim$200 \AA} with 
monatomic steps aligned parallel to the [1$\bar{1}\bar{1}$] direction. The second was a vicinal surface with a 
\mbox{4.6$^{\circ}$} miscut from the (110) plane, corresponding to an average terrace width of \mbox{$\sim$25 \AA} with 
monatomic steps oriented perpendicular to the [1$\bar{1}\bar{1}$] crystallographic direction. The film deposition and 
analysis were performed in a multi-chamber ultra-high vacuum (UHV) system with a base pressure in the low 
\mbox{$10^{-10}$ torr}. The substrate was cleaned by annealing in \mbox{$5 \times {10^{-7}}$ torr} O$_{2}$ at 
temperatures in the \mbox{$1300 \leq \textrm{T} \leq 1500$ K} range for \mbox{30-60 min} cycles, followed by 
flash-annealing several times to \mbox{2400 K} for \mbox{10-15 s} intervals in UHV. The Fe films were deposited by 
electron beam evaporation of a 3N purity Fe rod. The chamber pressure typically remained below \mbox{$2 
\times{10^{-10}}$ torr} during deposition. A quartz crystal balance was used to monitor the deposition rate, which 
typically lay between 0.006 and \mbox{0.03 \AA s$^{-1}$}. The deposition stage was equipped with a resistive heater and 
a k-type thermocouple to allow deposition on substrates at elevated temperatures. The STM results were obtained in 
constant current mode using a home-built instrument with either W or MnNi tips \cite{ceb03}. 

\section{Results and discussion}

The mechanisms guiding the formation of these nanowedge structures can be separated into three distinct terms: (1) the 
initial stages of nanowedge formation from nanowires produced in the first and second Fe layers, (2) the subsequent 
migration of adatoms towards the nanowedges and (3) the adatom migration within the thickening wedge.

\subsection{Initial stages of nanowedge formation}

The first Fe layer and up to \mbox{$\sim$80 \%} of the second layer form nanowires on Mo(110) by the step-flow growth 
mechanism in the \mbox{$495 \leq \textrm{T} \leq 525$ K} temperature range. The intermixing of Fe with the Mo(110) 
surface can be neglected for substrate temperatures up to \mbox{800 K} \cite{tik90}. Despite the large lattice mismatch, 
the first Fe layer grows pseudomorphically on Mo(110) due to the difference in surface energies between molybdenum 
\mbox{($\gamma_{Mo} = 2.95$ J.m$^{-2}$)} and iron \mbox{($\gamma_{Fe} = 2.55$ J.m$^{-2}$)} \cite{mez82}, which 
compensates for the large tensile strain accommodated by this layer. However, the increase in elastic energy upon 
addition of a second Fe layer creates favourable conditions for the formation of strain-relieving dislocations. 
Dislocation lines are formed along the [00$\bar{1}$] direction in second layer nanowires above a critical wire width of 
the order of \mbox{100 \AA} \cite{mur02}. The dislocations appear to be randomly distributed in wires that are 
\mbox{130--200 \AA} wide, but an array of closely-spaced dislocations is found in wires that are \mbox{300--600 \AA} 
wide \cite{mur03a}. 

\begin{figure}
\center
\includegraphics[width=0.6\textwidth,height=0.55\textheight]{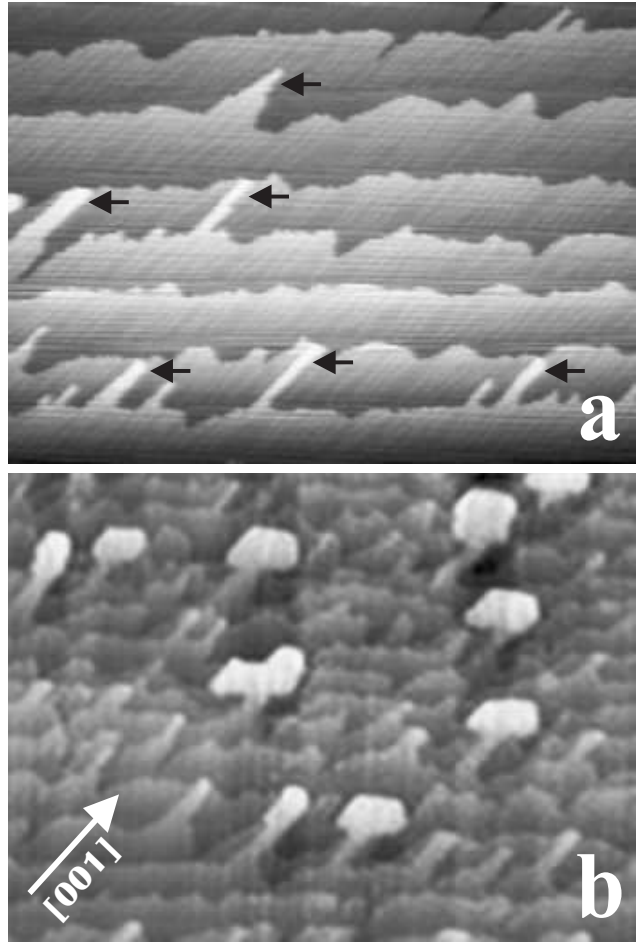}
\caption{\mbox{500 nm $\times$ 350 nm} STM images of a film with \mbox{3 \AA} nominal thickness grown on the low-index 
Mo(110) surface at \mbox{$495 \pm 15$ K}. (a) Fe third layer protrusions (marked with arrows) are formed on top of some 
of the dislocation lines in the second Fe layer. (b) Protrusions on successive terraces overlap and coalesce to form 
small nanowedges.}
\label{figure02}
\end{figure}

In \mbox{Fig. \ref{figure02}(a)} it is clear that third layer protrusions are formed along some of these dislocation 
lines. This reflects the preferential nucleation of third layer Fe islands at dislocations in the second Fe layer, which 
has been observed for Fe films grown on Mo(110) and also on W(110) near room temperature \cite{mur02,jen96}. The 
preferential nucleation is driven by the variation in lattice strain at the dislocation, where the insertion of extra Fe 
atoms leads to a local compressive strain. The protrusions shown in \mbox{Fig. \ref{figure02}(a)} are the starting point 
for nanowedge formation. When protrusions on successive terraces overlap and coalesce, small nanowedges are formed, as 
shown in \mbox{Fig. \ref{figure02}(b)}. This is accompanied by a roughening of the step-edges as the second layer 
nanowires break up and Fe atoms are absorbed into the developing nanowedge islands. 

\subsection{Adatom migration towards the nanowedge}

Both the presence of surface steps and the two-fold rotational symmetry of Mo(110) give rise to anisotropies in the 
migration of adatoms on this surface. However, we will show here that the localized strain caused on the surface by the 
Fe islands governs the migration of adatoms towards the nanowedges. 
 
Although much attention has been paid to the strain in a film caused by a substrate, relatively little attention has 
been paid to its inevitable consequence - strain in a substrate caused by a film, though the latter can be considerable 
\cite{san96}. To illustrate this point we performed a calculation, using the finite element method, on the simple model 
of a nanoscale Fe disk bonded to a Mo substrate. The thickness of the disk is \mbox{1 nm} and its diameter is in the 
range of 10 to \mbox{100 nm}. The thickness and the size of the substrate are much greater, so that the substrate could 
be considered as infinitely large when compared to the disk. The Fe disk is isotropically strained in the radial 
direction to reflect the lattice mismatch between the disk and the substrate \mbox{($\sim 9$ \%)}. The pattern of strain 
introduced in our model could be formulated as follows: a free non-supported disk is stretched uniformly so that the 
isotropic strain $\epsilon_{0}$ is induced in it and in this expanded state it is bonded to the substrate. The disk 
attempts to shrink but is held in tensile strain by the substrate. This introduces a corresponding compressive strain 
into the substrate beneath the disk. 

In the axisymmetric case of an isotropic medium, the strain $\epsilon$ and stress $\sigma$ vectors are linked with each 
other as follows:
\begin{equation}
\sigma = \textrm{C}(\epsilon - \epsilon_{0})
\end{equation}
Here $\epsilon_{0}$ stands for the initial strain of the disk to represent the lattice mismatch with the substrate. The 
strain vector has only four independent components
\begin{equation}
\pmatrix{\epsilon_{r}\cr
\epsilon_{\theta}\cr
\epsilon_{z}\cr
\gamma_{rz}\cr} = \pmatrix{\frac{\partial{u}}{\partial{r}}\cr
\frac{u}{r}\cr
\frac{\partial{w}}{\partial{r}}\cr
\frac{\partial{u}}{\partial{z}} + \frac{\partial{w}}{\partial{r}}}
\end{equation}
and not six as in the general non-symmetric case. $r, z, \theta$ are the coordinates of the cylindrical system. $u$ and 
$w$ are the radial and axial displacements respectively and are both independent of $\theta$. The tensor
\begin{equation}
\textrm{C} = \frac{E}{(1 + \nu)(1 - 2\nu)} \pmatrix{1 - \nu & \nu & \nu & 0\cr
\nu & 1 - \nu & \nu & 0\cr
\nu & \nu & 1 - \nu & 0\cr
0 & 0 & 0 & \frac{1 - 2\nu}{2}}
\end{equation}
includes the Youngs' modulus $E$ for the isotropic material and the Poisson's ratio $\nu$. We then use the finite 
element method to find the stress and strain in the disk and the substrate.

Here we use the values for the Youngs' modulae $E$ of \mbox{$211 \times{10^{9}}$} (SI Units) and \mbox{$329 
\times{10^{9}}$} (SI Units) and Poisson's ratios $\nu$ of 0.33 and 0.293 for Fe and Mo respectively. One can appreciate 
that based on the continuum approximation this model has its limitations, in particular when dealing with objects as 
small as tens of nanometers in size. Besides, one can appreciate that as the films grown are epitaxial, the modulus of 
elasticity $E$ has a number of components (in the case of a cubic crystal it contains three components) \cite{lan70}. 
However, as this calculation aims at a qualitative rather than quantitative result, we will avoid complications in the 
model by considering it to be isotropic. Another issue is the elasticity limit; our model assumes that all the 
deformations remain elastic. In reality, the elasticity limit is exceeded due to the large lattice mismatch between Fe 
and Mo. Finally, the model does not take into account that the nanowedge islands grow on top of a single pseudomorphic 
Fe wetting layer, as is the case in the experiment \cite{mur02}.

\begin{figure}
\center
\includegraphics[width=0.5\textwidth,height=0.5\textheight]{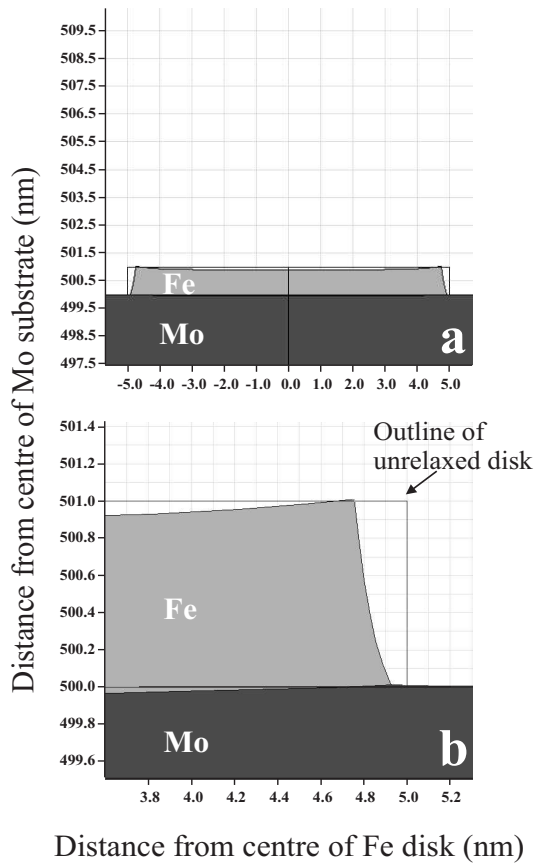}
\caption{(a) Simulated pattern of deformation in the strained Fe disk and in the Mo substrate. A \mbox{10 nm} diameter 
Fe disk (\mbox{1 nm} thick) is strained on top of a Mo substrate, \mbox{2000 nm} in diameter and \mbox{1000 nm} thick, 
and allowed to relax. (b) A zoom-in of the edge of the Fe disk, where the deformation of the disk is evident from the 
outline of the disk before it is allowed to relax. A local deformation is induced in the substrate near the edge of the 
disk because the substrate is stretched inwards along the radial direction of relaxation of the disk.}
\label{figure03}
\end{figure}

\begin{figure}
\center
\includegraphics[width=0.6\textwidth,height=0.65\textheight]{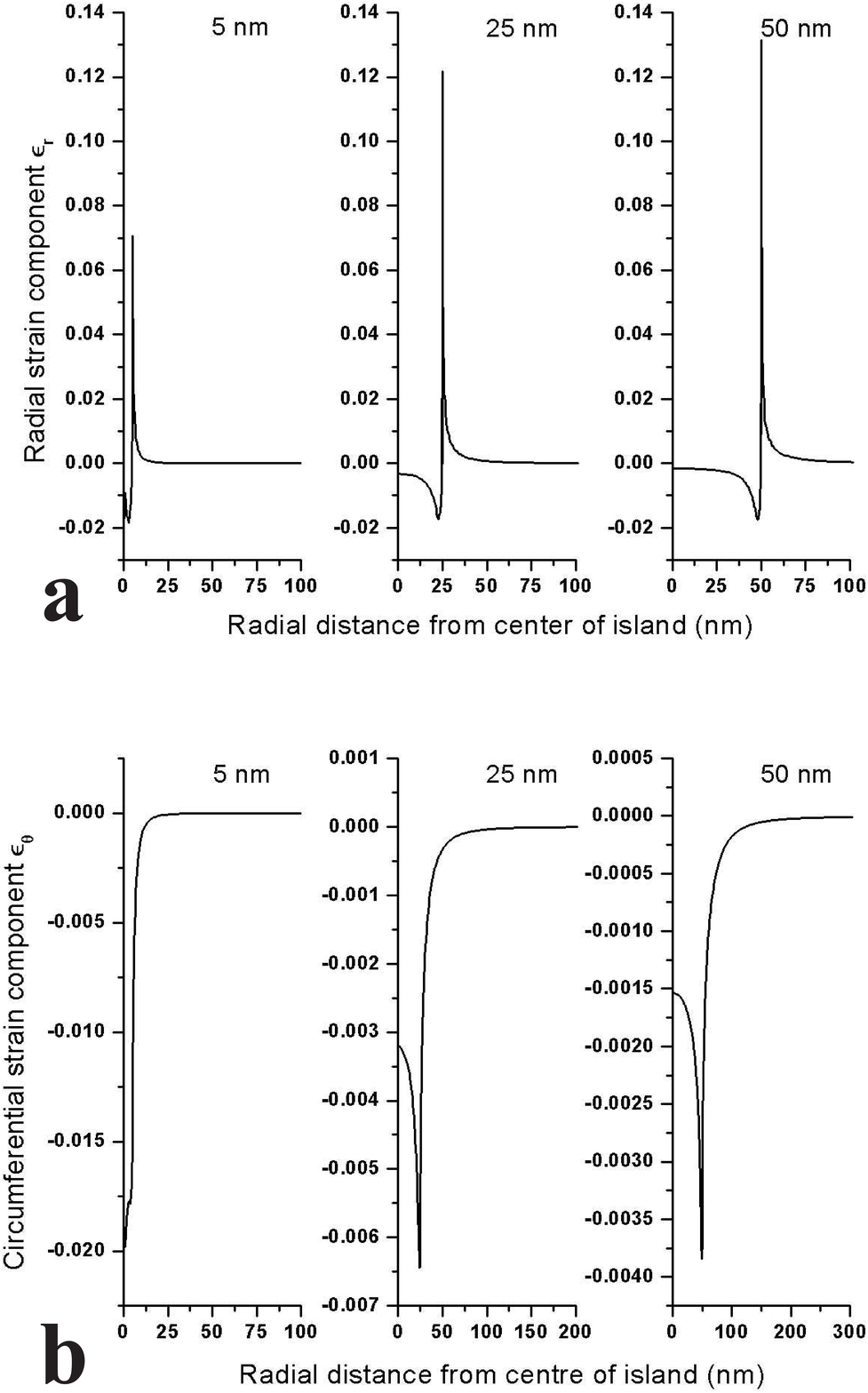}
\caption{Calculated (a) radial and (b) circumferential strain profiles for a Mo substrate covered by \mbox{1 nm} thick 
Fe disks of 5, 25 and \mbox{50 nm} radii, respectively. (a) The substrate displays a compressive radial strain 
underneath the Fe disk and a tensile strain in the area outside the disk edge. The radial strain falls off to \mbox{1 
\%} of its maximum value at approximately 9, 18 and \mbox{22 nm} from the edge of the 10, 50 and \mbox{100 nm} diameter 
disks, respectively. (b) The circumferential strain in the substrate outside each Fe disk edge is compressive, but is 
smaller in magnitude than the radial strain.}
\label{figure04}
\end{figure}

Figures \ref{figure03}(a) and (b) show the simulated pattern of deformation in a cross-section perpendicular to the 
surface at the edge of the disk, resulting from the model. In the process of relaxing the tensile strain in the disk, 
the substrate near the edge of the disk is stretched inwards along the radial direction of relaxation, producing a 
tensile strain in the substrate around the disk edge. The relaxation also introduces a compressive circumferential 
strain into the substrate outside the disk edge. \mbox{Figure \ref{figure04}} shows the dependencies of the radial 
strain component $\epsilon_{r} = \partial{u}/\partial{r}$ and circumferential strain component $\epsilon_{\theta} = u/r$ 
at the surface of the substrate as a function of the distance from the centre of the disk. As expected intuitively, the 
strain falls off more slowly from the disk edge in larger diameter disks, indicating that islands of greater size create 
larger strained areas in the substrate. 

The results of this simple model agree qualitatively with recent atomic scale calculations for Co islands on Cu(111) 
\cite{tsi03}. These calculations have shown that the Co islands induce an anisotropic strain in the Cu substrate in 
their vicinity and furthermore, that this strain influences the adatom migration barrier such that adatom motion 
parallel to an island edge is preferred to adatom motion perpendicular to it. This is due to the fact that for 
transition metals, the adatom migration barrier generally increases with the surface tensile strain \cite{sab03}, which 
is reflected by a preference for nucleation on these areas. However, this is not the case for the Fe/Mo(110) system, 
where we have already observed by STM that Fe adatoms prefer nucleation on top of dislocations where there is a local 
compressive strain. 

The effect of surface strain on the diffusion and nucleation of adatoms must be considered in terms of both the adatom 
binding energy and the diffusion barrier. In the case of a small tensile strain (\mbox{1--3 \%}), the interlayer 
separation between the first and second layers of the surface is reduced. Consequently, an adatom occupying a hollow 
site on a surface will experience greater binding to the subsurface atom directly beneath it. This increases the 
diffusion barrier to an adjacent hollow site. The situation is reversed for a compressively strained surface, so that 
the adatom experiences a lowered diffusion barrier. However, in the case of a large tensile strain (\mbox{$>$3 \%}), the 
energy of the bridge site, which the adatom temporarily occupies as it hops between adjacent hollow sites, is also 
lowered, causing the diffusion barrier to decrease again \cite{bru95}. This siutation can occur in the Fe/Mo(110) 
system, where the pseudomorphic Fe film must accommodate \mbox{$\sim$9 \%} tensile mismatch strain. In this situation, 
the diffusion barrier across a dislocation may be higher than the diffusion barrier in the tensile regions between the 
dislocations.    

\begin{figure}
\center
\includegraphics[width=0.6\textwidth,height=0.3\textheight]{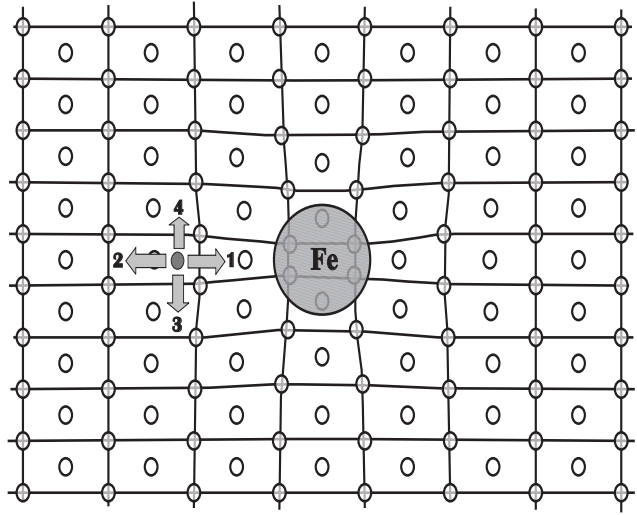}
\caption{A schematic representation of the tensile strain field produced in the surface around an Fe island. The open 
circles represent the (110) surface atoms - the surface net is also shown. The large shaded circle is an Fe island, 
while the dark circle is a single diffusing Fe adatom in a quasi threefold site. The four possible migration directions 
are indicated.}
\label{figure05}
\end{figure}

The migration of Fe adatoms towards the nanowedge islands can now be explained in terms of \mbox{Fig. \ref{figure05}}, 
which illustrates the case of an Fe adatom located on the pseudomorphic Fe wetting layer in the vicinity of an island. A 
pattern of enhanced tensile strain (\mbox{$\geq$9 \%}) is produced in the surface, which is roughly radial with the 
center located at the island. Adatom migration along direction 1 leads to a decrease in the adatom diffusion barrier, 
while direction 2 leads to its increase and directions 3 and 4 are energetically equivalent. The net result is an 
anisotropic migration of adatoms along the direction of radial strain towards the Fe islands. As the size of the 
strained area is dependent on the size of the island, the separation between the islands increases with their size, 
which is typically what is observed by STM. It is also expected that the adatom migration will be directed towards the 
thicker end of each nanowedge, which produces a greater strain in the surface. 

\subsection{Adatom migration within the nanowedge}

\begin{figure}
\center
\includegraphics[width=0.4\textwidth,height=0.7\textheight]{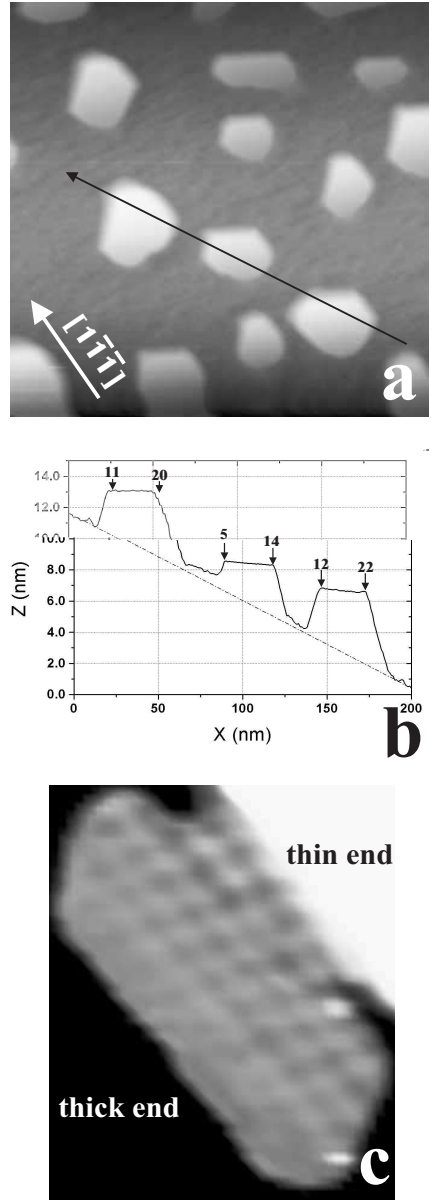}
\caption{(a) 200 nm $\times$ 200 nm STM image of an iron film of \mbox{6 \AA} nominal thickness grown on vicinal Mo(110) 
at \mbox{$700 \pm 25$ K}. The black arrow marks the position of the line profile shown in (b), which was taken from the 
thin end of the nanowedges to their thick end. (b) The local thickness at the thin and thick ends of each nanowedge is 
indicated by the number of Fe layers. (c) Example of an iron nanowedge where the corrugation produced by the dislocation 
network decreases towards the thicker end of the island.}
\label{figure06}
\end{figure}

\begin{figure}
\center
\includegraphics[width=0.65\textwidth,height=0.6\textheight]{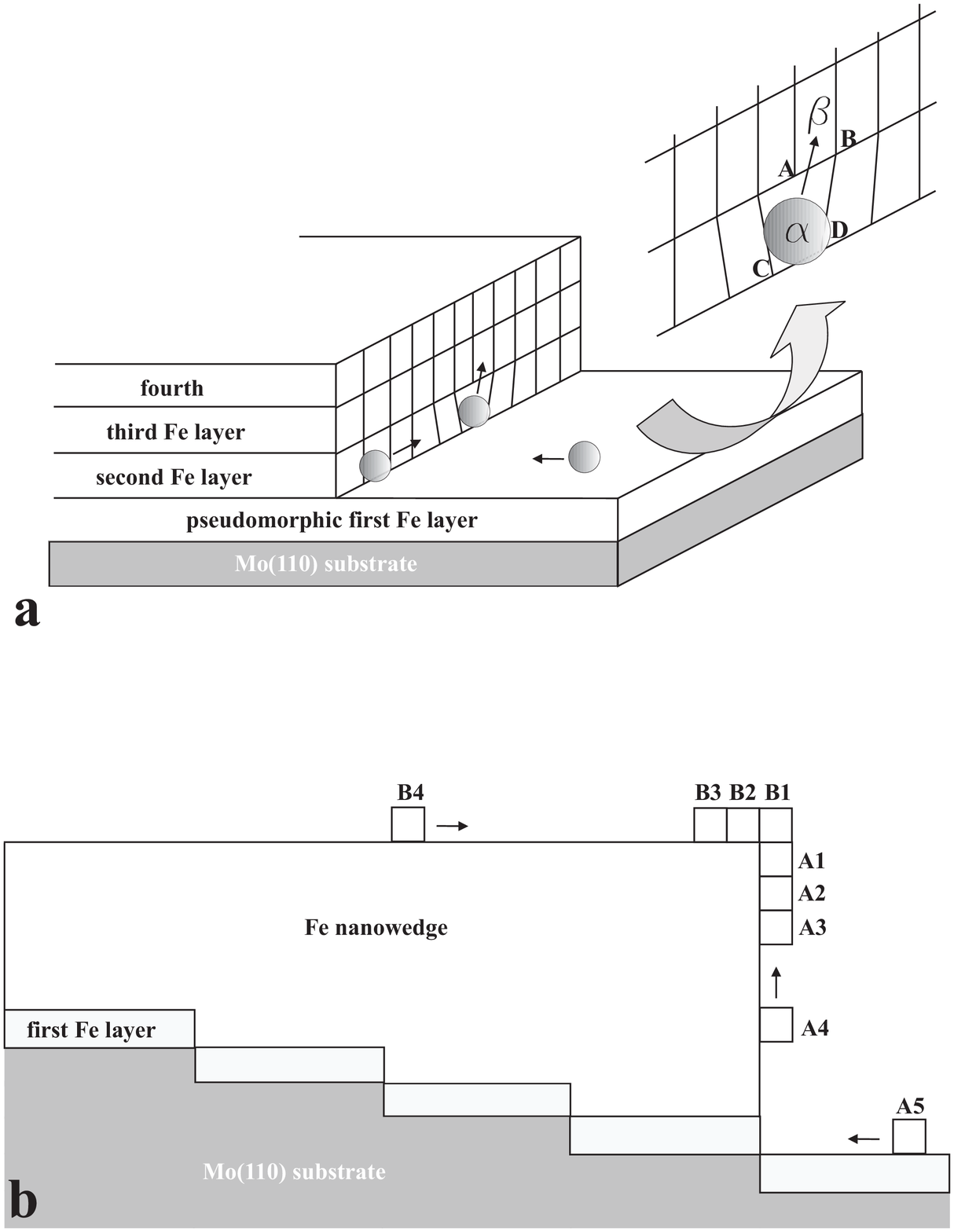}
\caption{(a) Atoms migrate across the pseudomorphic first Fe layer until they reach the edge of the wedge. They migrate 
along this edge until they reach the position of a dislocation - this region is enlarged in the insert. Here the 
interatomic separation AB in the third Fe layer is smaller than the separation CD in the second layer. The atom can 
migrate from position $\alpha$ to position $\beta$ despite the energy increase caused by the reduction in coordination, 
as it is compensated for by the increased adatom binding in position $\beta$. (b) Schematic model showing the migration 
of atoms (A1-A4) up the thick face of a nanowedge to form an overhanging structure. Atoms deposited on the surface of 
the nanowedge (B1-B4) also migrate towards the thick end of the nanowedge.}
\label{figure07}
\end{figure}

From the STM evidence shown in \mbox{Fig. \ref{figure06}(a)-(b)}, it is clear that depending on the nominal film 
thickness deposited, the nanowedges can be many Fe layers thick at their {\em thin} end. Therefore, to explain the later 
stages of the nanowedge formation, we must adopt a model where the adatoms undertake a vertical climb up each side of 
the nanowedge. We propose that dislocations in the lower layers of the nanowedges act as migration channels to 
facilitate this vertical climb model. Dislocation lines and networks are formed in the lower layers of each nanowedge 
\cite{mal98,mur02}, which thread to the walls of the island. Climbing up an island wall implies a reduction in the 
coordination of the adatom as it makes the very first upward step because it loses contact with atoms on the substrate. 
However, by considering \mbox{(Fig. \ref{figure07}(a))}, our model can explain such vertical transport. If the adatom 
moves from the position $\alpha$ to the position $\beta$, thereby making the first step up along the vertical wall, it 
loses contact with the nearest neighbours on the substrate and this increases the energy of the adatom. Following the 
model of adatom binding energy dependent on the separation between the adatom and its nearest neighbours, we can 
appreciate that the required reduction in energy of the adatom comes from the fact that the separation between the atoms 
AB is smaller than the one between the atoms CD, due to the presence of the dislocation. The strongest change in the 
interatomic separation happens in the immediate vicinity of the dislocation. Therefore, the locations where the 
dislocation lines exit to the side of the island become the channels for the vertical climb of adatoms upwards along the 
island walls.   

The introduction of dislocations within the lower layers of the nanowedge produces a sharp reduction in the tensile 
strain over the first few layers. As a result, the lattice strain within these layers corresponds to the small tensile 
strain (\mbox{1--3 \%}) regime discussed earlier, where the adatom experiences an increased diffusion barrier. The 
further up the adatom climbs along the vertical wall of the island, the more this tensile strain is reduced, as the 
interatomic separation approaches the bulk Fe(110) value within the thick end of the nanowedge. This is illustrated in 
\mbox{Fig. \ref{figure06}(c)}, where the surface corrugation of the dislocation network decreases with increasing island 
thickness because the Fe lattice becomes relaxed. The reduction in tensile strain towards the thick end of the island 
lends itself to anisotropic surface migration in this direction. This is schematically shown in \mbox{Fig. 
\ref{figure07}(b)}. Adatoms A1, A2, A3 have climbed the wall of the island to the point that is most remote from the 
substrate where the interatomic separation between the atoms of the wall is closest to the bulk value. Adatom A4 is 
shown schematically moving up the wall. We could therefore suggest that the walls of the nanowedges can form hanging 
cliffs with negative slope. Unfortunately, these cannot be readily observed using STM. Likewise, adatoms B1, B2, B3 and 
B4 deposited on the surface of the island, migrate towards the thick end of the wedge for the same reason. The net 
effect is that the nanowedge islands continue to develop predominantly from their thick end.

\section{Conclusions}

We have experimentally observed an interesting film growth mode using the Fe/Mo(110) epitaxial system. The film forms 
nanowedges, i.e. structures that can be down to one or two monolayers thick at one end and many monolayers thick at the 
other. The thickness of the nanowedges increases with each atomic terrace step of the substrate so that the top surface 
of the nanowedge is virtually atomically flat. This growth mode was also observed before in Fe/W(110) system. The growth 
of nanowedges implies a most peculiar adatom mass transport on the surface, which has not been explained before. 

We propose a model explaining the growth of these wedge-shaped islands. The model is based on the strain caused by the 
lattice mismatch between the film and the substrate. The essential aspect of the model is the strain caused by the 
nanoscale Fe islands within the substrate. Our model suggests that adatoms migrate towards the islands due to the 
influence of this strain. The model further suggests that adatoms enter the islands not only through their thin ends 
where the island is only two monolayers thick but on the contrary, through a vertical climb along the sides, which is 
driven again by the energy reduction through movement towards the location where the interatomic spacing corresponds to 
the bulk value.

One of the key elements of the model is that misfit dislocations are formed in the lower layers of the film. This is 
based on our experimental STM data. The model agrees qualitatively well with our experimental results. However, it is 
clear that there are aspects of the mass transport that require a more refined model. We expect that self-assembly of 
these kinds of nanowedges could take place in other systems characterised by large misfit between the film and the 
substrate. These nanostructures could also have interesting electronic and magnetic properties.

\section{Acknowledgments}

This work was supported by the Science Foundation Ireland under contract 00/PI.1/C042.

\newpage

\newpage

\newpage

\newpage

\newpage

\newpage

\newpage

\begin{thebibliography}{22}

\bibitem{bru98}
H. Brune, M. Giovannini, K. Bromann, K. Kern,
\newblock Nature 394 (1998) 451.

\bibitem{bru02}
K. Brunner,
\newblock Rep. Prog. Phys. 65 (2002) 27.

\bibitem{tei02}
C. Teichert,
\newblock Phys. Rep. (2002) 335.

\bibitem{ale98}
M. Alexe, J.F. Scott, C. Curran, N.D. Zakharov, D. Hesse, A. Pignolet,
\newblock Appl. Phys. Lett. 73 (1998) 1592.

\bibitem{bat01}
M. Batzill, D.E. Beck, B.E. Koel,
\newblock Appl. Phys. Lett. 78 (2001) 2766.

\bibitem{vas03}
E. Vasco, R. Dittman, S. Karth\"auser, R. Waser,
\newblock Appl. Phys. Lett. 82 (2003) 2497. 

\bibitem{mal98}
J. Malzbender, M. Pryzbylski, J. Giergiel, J. Kirschner,
\newblock Surf. Sci. 414 (1998) 187.

\bibitem{mur02}
S. Murphy, D. Mac Math\'una, G. Mariotto, I.V. Shvets,
\newblock Phys. Rev. B 66 (2002) 195417.

\bibitem{bet95}
H. Bethge, D. Heuer, C. Jensen, K. Besh\"oft, U. K\"ohler,
\newblock Surf. Sci. 331-333 (1995) 878.

\bibitem{res99}
K. Resh\"oft, C. Jensen, U. K\"ohler,
\newblock Surf. Sci. 421 (1999) 320.

\bibitem{sch98}
C. Schmidthals, D. Sander, A. Enders, J. Kirschner,
\newblock Surf. Sci. 417 (1998) 361.

\bibitem{pia00}
G. Piaszenski, R. G\"obel, C. Jensen, U. K\"ohler,
\newblock Surf. Sci. 454-456 (2000) 712.

\bibitem{alt97}
I.B. Altfeder, K.A. Matveev, D.M. Chen,
\newblock Phys. Rev. Lett. 78 (1997) 2815.

\bibitem{ceb03}
S.F. Ceballos, G. Mariotto, S. Murphy, I.V. Shvets,
\newblock Surf. Sci. 523 (2003) 131.

\bibitem{tik90}
M. Tikhov, E. Bauer,
\newblock Surf. Sci. 232 (1990) 73.

\bibitem{mez82}
L. Mezey, J. Giber,
\newblock Surf. Sci. 117 (1982) 220.

\bibitem{mur03a}
S. Murphy, J. Osing, I.V. Shvets,
\newblock Surf. Sci. 547 (2003) 139.

\bibitem{jen96}
C. Jensen, K. Resh\"oft, U. K\"ohler,
\newblock Appl. Phys. A 62 (1996) 217.

\bibitem{san96}
D. Sander, R. Skomski, C. Schmidthals, A. Enders, J. Kirschner,
\newblock Phys. Rev. Lett. 77 (1996) 2566.

\bibitem{lan70}
L.D. Landau, E.M. Lifshitz,
\newblock Theory of Elasticity, 2nd ed. (Pergammon Press, Oxford, 1970).  

\bibitem{tsi03}
D.V. Tsivlin, V.S. Stepanyuk, W. Hergert, J. Kirschner,
\newblock Phys. Rev. B 68 (2003) 205411.

\bibitem{sab03}
R.F. Sabiryanov, M.I. Larsson, K.J. Cho, W.D. Nix, B.M. Clemens,
\newblock Phys. Rev. B 67 (2003) 125412.

\bibitem{bru95}
H. Brune, K. Bromann, H. R\"oder, K. Kern, J. Jacobsen, P. Stoltze, K. Jacobsen, J. N{\o}rskov,
\newblock Phys. Rev. B 52 (1995) R14380. 

\end{thebibliography}
\end{document}